\documentclass[aip,jap,reprint,natbib,showkeys,superscriptaddress] {revtex4-1}

\usepackage{graphicx}
\usepackage{dcolumn}
\usepackage{amsmath}

\DeclareMathOperator{\sgn}{sgn}
\DeclareMathOperator{\Hev}{H}

\begin{document}

\title{Adiabatic Superconducting Artificial Neural Network: Basic Cells}

\author{Igor I. Soloviev}
\email[]{isol@phys.msu.ru} \affiliation{Lomonosov Moscow State
University Skobeltsyn Institute of Nuclear Physics, 119991 Moscow,
Russia} \affiliation{MIREA - Russian Technological University,
119454, Moscow, Russia} \affiliation{Moscow Institute of Physics and
Technology, 141700 Dolgoprudny, Russia}

\author{Andrey E. Schegolev}
\affiliation{Lomonosov Moscow State University Skobeltsyn Institute
of Nuclear Physics, 119991 Moscow, Russia} \affiliation{MIREA -
Russian Technological University, 119454, Moscow, Russia}
\affiliation{Physics Department, Lomonosov Moscow State University,
119991 Moscow, Russia}\affiliation{Moscow Technical University of
Communications and Informatics (MTUCI), 111024, Moscow, Russia}

\author{Nikolay V. Klenov}
\affiliation{Lomonosov Moscow State University Skobeltsyn Institute
of Nuclear Physics, 119991 Moscow, Russia} \affiliation{MIREA -
Russian Technological University, 119454, Moscow, Russia}
 \affiliation{Physics Department, Lomonosov Moscow State University, 119991
Moscow, Russia}\affiliation{Moscow Technical University of
Communications and Informatics (MTUCI), 111024, Moscow, Russia}

\author{Sergey V. Bakurskiy}
\affiliation{Lomonosov Moscow State University Skobeltsyn Institute
of Nuclear Physics, 119991 Moscow, Russia} \affiliation{MIREA -
Russian Technological University, 119454, Moscow, Russia}
\affiliation{Moscow Institute of Physics and Technology, 141700
Dolgoprudny, Russia}

\author{Mikhail Yu. Kupriyanov}
\affiliation{Lomonosov Moscow State University Skobeltsyn Institute
of Nuclear Physics, 119991 Moscow, Russia}

\author{Maxim V. Tereshonok}
\affiliation{MIREA - Russian Technological University, 119454,
Moscow, Russia} \affiliation{Moscow Technical University of
Communications and Informatics (MTUCI), 111024, Moscow, Russia}

\author{Anton V. Shadrin}
\affiliation{Moscow Institute of Physics and Technology, 141700
Dolgoprudny, Russia}

\author{Vasily S. Stolyarov}
\affiliation{Moscow Institute of Physics and Technology, 141700
Dolgoprudny, Russia} \affiliation{Institute of Solid State Physics
RAS, 142432 Chernogolovka, Russia} \affiliation{Fundamental Physical
and Chemical Engineering dep., MSU, 119991 Moscow, Russia}
\affiliation{Solid State Physics Department, KFU, 420008 Kazan,
Russia}

\author{Alexander A. Golubov}
\affiliation{Moscow Institute of Physics and Technology, 141700
Dolgoprudny, Russia} \affiliation{Faculty of Science and Technology
and MESA+ Institute of Nanotechnology, 7500 AE Enschede, The
Netherlands}


\begin{abstract}
We consider adiabatic superconducting cells operating as an
artificial neuron and synapse of a multilayer perceptron (MLP).
Their compact circuits contain just one and two Josephson junctions,
respectively. While the signal is represented as magnetic flux, the
proposed cells are inherently nonlinear and close-to-linear magnetic
flux transformers. The neuron is capable of providing a one-shot
calculation of sigmoid and hyperbolic tangent activation functions
most commonly used in MLP. The synapse features by both positive and
negative signal transfer coefficients in the range $\sim
(-0.5,0.5)$. We briefly discuss implementation issues and further
steps toward multilayer adiabatic superconducting artificial neural
network which promises to be a compact and the most energy-efficient
implementation of MLP.
\end{abstract}



\maketitle

\section{Introduction}

Artificial neural network (ANN) is the key technology in the fast
developing area of artificial intelligence. It has been already
broadly introduced in our everyday life. Further progress requires
an increase in complexity and depth of ANNs. However, modern
implementations of neural networks are commonly based on
conventional computer hardware which does not suit well for
neuromorphic operation. This leads to excessive power consumption
and hardware overhead. Ideal basic elements of ANNs should combine
the multiple properties like one-shot calculation of their
functions, operation with energy near the thermal noise floor and
nanoscale dimensions.

The most energy efficient computing today can be performed using the
superconductor digital technology \cite{Beil}. The first ever
practical logic gates capable of operating down to and below the
Landauer thermal limit \cite{Land} were realized recently
\cite{REVASL} on the basis of adiabatic superconductor logic.
Alongside the several attempts to implementation of the
superconducting ANNs proposed since 1990-s
\cite{ANN1,ANN2,ANN3,ANN4,ANN6,ANN7,ANN8,ANN9,ANN10}, the idea to
adopt the adiabatic logic cells to neuromorphic circuits was
presented only recently \cite{BeilANN,{IEEEANN}}. In this paper, we
consider operation principles of adiabatic superconducting basic
cells which comply with the above-mentioned properties for ANN
implementation. We focus on a particular multilayer perceptron (MLP)
because of a wide range of its applicability and well-developed
learning algorithms for such a network.

\section{Basic cells}

The basic element of superconducting circuits is the Josephson
junction. Its characteristic energy typically lies below aJ level
while switching frequency is several hundreds GHz. Contrary to
semiconductor transistor, Josephson junction is not fabricated in a
substrate but between two superconductor layers deposited on a
substrate utilized as a mechanical support. This provides
opportunity for superconducting circuits to benefit from 3D topology
which can be especially suitable for deep ANNs. The minimal feature
size of superconducting circuits is progressively decreased down to
nanoscales in recent years \cite{Fab}.

Another attractive feature of Josephson junction is its inherently
strong nonlinearity. Indeed, the current flowing through the
junction, $I$, is commonly related to the superconducting phase
difference between the superconducting banks, $\varphi$, as
\begin{equation}
I = I_c\sin\varphi, \label{CPR}
\end{equation}
where $I_c$ is the junction critical current. We show below that
this current-phase relation (CPR) having both linear and nonlinear
parts is well suited for implementation of superconducting
artificial neuron with one-shot calculation of sigmoid or hyperbolic
tangent activation functions,
\begin{subequations} \label{SigTaf}
\begin{equation}
\sigma(x) = \frac{1}{1+e^{-x}}
\end{equation}
\begin{equation}
\text{or} \nonumber
\end{equation}
\begin{equation}
\tau(x) = \tanh(x),
\end{equation}
\end{subequations}
utilized in MLP, and superconducting synapse enabling signal
transfer with both positive and negative coefficients. Unlike most
of their predecessors
\cite{ANN1,ANN2,ANN3,ANN4,ANN6,ANN7,ANN9,ANN10} both cells are
operating in a pure superconducting mode featured by minimal power
consumption.

\subsection{Artificial neuron}

One of the simplest superconducting cells is parametric quantron
proposed in 1982 for adiabatic operation \cite{PQ}. It is the
superconducting loop consisted of a Josephson junction and a
superconducting inductance. According to Josephson junction CPR
(\ref{CPR}), the relation between input magnetic flux and Josephson
junction phase in its circuit has a simple expression:
\begin{equation} \label{PQeq}
\varphi + l\sin\varphi = \phi_{in},
\end{equation}
where we use normalization of current to critical current of
Josephson junction, $I_c$, and input magnetic flux $\Phi_{in}$ to
the magnetic flux quantum $\Phi_0$, $\phi_{in} =
2\pi\Phi_{in}/\Phi_0$, inductance, L, is normalized to
characteristic inductance, $l = L/L_c$, $L_c = \Phi_0/2\pi I_c$,
accordingly.

It is seen from (\ref{CPR}), (\ref{PQeq}) that the current
circulating in the loop has a tilted sine dependence on input
magnetic flux. The way to transform this dependence close to the
desired one ((\ref{SigTaf}a) or (\ref{SigTaf}b)) is the addition of
a linear term compensating the sine slope on the initial section
(where $\sin\varphi \approx \varphi$) in the vicinity of zero input
flux, $\phi_{in} \approx 0$.
\begin{figure}[t]
\includegraphics[width=1\columnwidth,keepaspectratio]{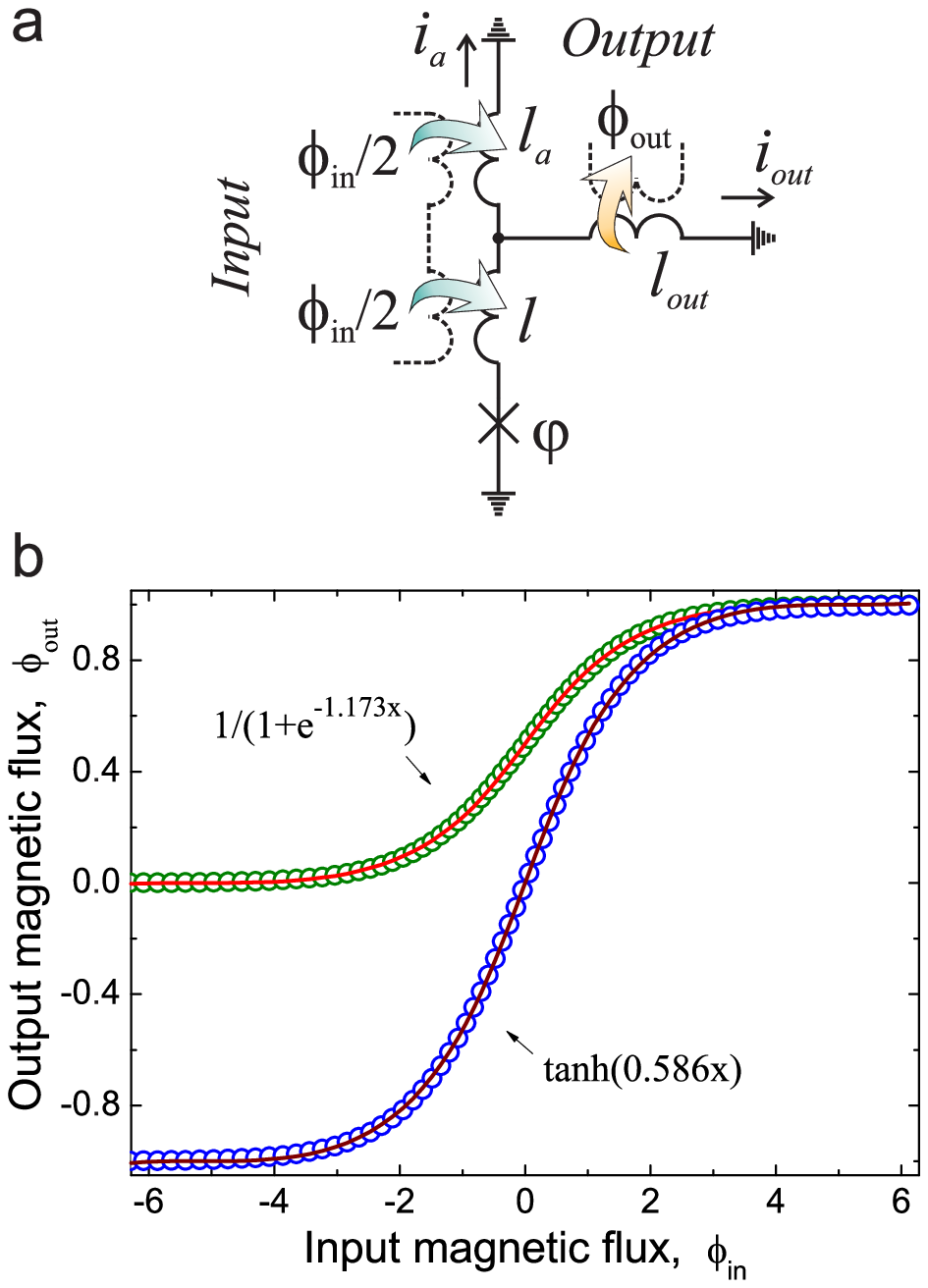}
\caption{(color online). (a) Scheme of an artificial neuron cell.
(b) The cell transfer function (line) fitted to sigmoid and
hyperbolic tangent functions (dots). Scaling of the functions
(\ref{SigTaf}) is shown in the figure. The transfer function
$\phi_{out}(\phi_{in})$ is normalized by $2\pi l_{out}/(l_a + 2
l_{out})$ and shifted by $-2 \pi (l_a + l_{out})/(l_a + 2 l_{out})$
on the flux axis to fit (\ref{SigTaf}a), and normalized to $\pi
l_{out}/(l_a + 2l_{out})$ with no additional shift on flux axis to
fit (\ref{SigTaf}b). The optimal values of parameters are $l =
0.125$, $l_{out} = 0.3$, $l_a = 1.125$. Consistency of curves in
both cases is at the level of $10^{-3}$. Hyperbolic tangent
activation function is fitted with $\pi$ shift in Josephson junction
CPR (\ref{CPR}).} \label{Fig1}
\end{figure}

This can be done by attaching another superconducting loop with a
part of its inductance, $l_{out}$, being common with the initial
circuit, see Fig.~\ref{Fig1}a. The synthesized cell was named a
``sigma-cell'' \cite{BeilANN} because its transformation of magnetic
flux can be very close to sigmoid function. Here we are interested
in a transfer function, $\phi_{out}(\phi_{in})$, where output
magnetic flux, $\phi_{out}$, is proportional to output current,
$\phi_{out} = l_{out} i_{out}$.

The system of equations describing the proposed cell is as follows:
\begin{subequations} \label{SCeq}
\begin{equation}
\varphi + l\sin\varphi = \phi_{in}/2 + l_{out} i_{out},
\end{equation}
\begin{equation}
\varphi + l\sin\varphi = \phi_{in} + l_a i_a,
\end{equation}
\end{subequations}
where $l_a$ is the attached inductance. The corresponding system
implicitly defining the transfer function through dependencies of
$\phi_{out}$, $\phi_{in}$ on $\varphi$ has the following form:
\begin{subequations} \label{iq_fe}
\begin{equation}
\phi_{out} = l_{out}\frac{\phi_{in} - 2 l_a \sin\varphi}{2(l_a +
l_{out})},
\end{equation}
\begin{equation}
\phi_{in} = 2 \left(\frac{l_a + l_{out}}{l_a + 2
l_{out}}\right)\left[\varphi + \left(l + \frac{l_a l_{out}}{l_a +
l_{out}}\right)\sin\varphi\right].
\end{equation}
\end{subequations}
Vanishing of the derivative $d\phi_{out}/d\phi_{in}$ at $\phi_{in} =
0$ corresponds to the condition:
\begin{equation} \label{la_l+1}
l_a = 1 + l.
\end{equation}
One can fit (\ref{iq_fe}) to sigmoid function (\ref{SigTaf}a) taking
(\ref{la_l+1}) into account with the two fitting parameters: $l,
l_{out}$.

The result of fitting is shown in Fig.~\ref{Fig1}b. The found
optimal values, $l = 0.125$, $l_{out} = 0.3$, provide conformity of
the sigma-cell transfer function with sigmoid one with standard
deviation at the level of $10^{-3}$. Sigmoid function
(\ref{SigTaf}a) was scaled as $\sigma(1.173 x)$ in our fitting
process. The transfer function $\phi_{out}(\phi_{in})$ (\ref{iq_fe})
was normalized by $2\pi l_{out}/(l_a + 2l_{out})$ to fit a unit
height, and shifted by a half period. The latter can be obtained by
application of a constant bias flux to the circuit, $\phi_b = - 2
\pi (l_a + l_{out})/(l_a + 2 l_{out})$.

While sigmoid activation function is commonly used for input data
defined in the positive domain, for data defined on the whole
numeric axis around zero it is convenient to use hyperbolic tangent.
Application of additional bias flux providing $\pi$ phase shift into
the loop containing Josephson junction moves the center of the
nonlinear part of the cell transfer function to zero. This allows
obtaining the desired shape of activation function (\ref{SigTaf}b).
The $\pi$ phase shift can be also implemented using the $\pi$ -
Josephson junction \cite{PiJJ,GKI,SIsFS2013PRB,SIsFS2013APL} with
$\pi$ shift of its CPR (\ref{CPR}),  $I = - I_c \sin(\varphi)$,
instead of the standard one.

One need to correspondingly change the sign of the terms containing
sine function in (\ref{iq_fe}) to perform the fitting procedure. The
fitting result is presented in Fig.~\ref{Fig1}b. Hyperbolic tangent
function was scaled as $\tanh(0.586 x)$ while the transfer function
$\phi_{out}(\phi_{in})$ was normalized by a factor of two lower
value than the previous time, $\pi l_{out}/(l_a + 2l_{out})$. With
the same values of parameters $l$, $l_{out}$ and zero bias flux we
obtained the same conformity of the curves.

\subsection{Artificial synapse}

Synapse modulates the ``weight'' of a signal arriving at the neuron.
In our case the signal corresponds to magnetic flux and therefore
synapse can be implemented simply as a transformer of magnetic flux
with desired coupling factor. Summation of signals can be provided
by connecting the transformers to a single superconducting input
loop of the neuron. However, this solution suits for ANN with a
certain and unchangeable configuration.

In the most cases a configurable ANN would be preferable. The
selected configuration of inter-neuron connections should be
maintained during its entire use if the feature space dimensions do
not vary. However, the weights values should be configurable if we
want to train the ANN on the fly. The best way to meet this
requirement is utilization of some non-volatile memory elements. In
superconducting circuits such element can be implemented by using
the ferromagnetic (F) materials. In particular, introduction of
F-layers into Josephson junction weak link area allows to modulate
its critical current \cite{Beil,MJJ1,MJJ2}. This phenomenon was
already proposed for utilization in artificial synapse of
superconducting spiking ANN \cite{ANN10}. In our case of MLP we can
also make use of it.

\begin{figure}[t]
\includegraphics[width=1\columnwidth,keepaspectratio]{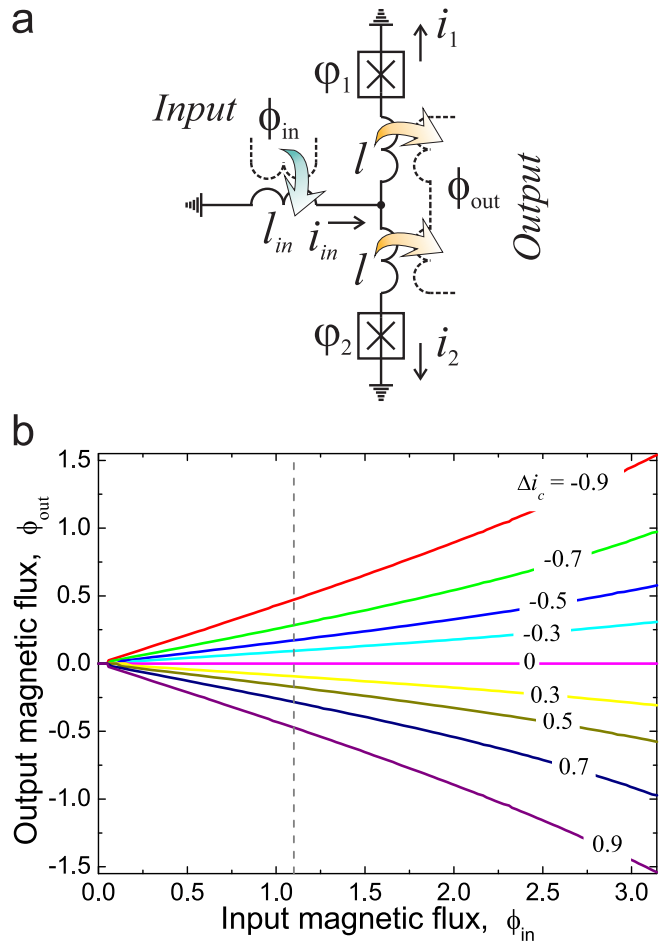}
\caption{(color online). (a) Scheme of an artificial synapse cell.
Magnetic Josephson junctions are marked by boxes. (b) Synapse cell
transfer function for the values of parameters: $l_{in} = 2$, $l =
4$, $\Sigma i_c = 1$ and $\Delta i_c$ as shown in the figure.
Vertical dotted line shows the boundary of highly linear range where
standard deviation from the linear function is at the level of
$10^{-3}$. This range corresponds to maximum output magnetic flux of
the optimized neuron cell.} \label{Fig2}
\end{figure}

The synapse scheme presented in Fig.~\ref{Fig2}a is nearly mirrored
scheme of the proposed neuron (Fig.~\ref{Fig1}a). The only
differences are the addition of the second Josephson junction and
the possibility to independently modulate critical currents of the
magnetic junctions (marked by boxes), e.g., by application of tuning
magnetic field.

For MLP it is required to provide both positive and negative weights
of signal. Our synapse is designed according to this requirement.
The input current, $i_{in}$, induced in inductance $l_{in}$ by input
magnetic flux, $\phi_{in}$, is split toward the two Josephson
junctions. Magnitude of currents $i_1$, $i_2$ in each branch
correspond to critical currents of the junctions, $i_{c1}$,
$i_{c2}$, so that the sign of output circulating current, $i_{cir} =
(i_1 - i_2)/2$, (and the direction of output magnetic flux,
$\phi_{out}$) is determined by their ratio. Maximum inequality of
$i_{c1}$, $i_{c2}$ provides maximum output signal, while equal
critical currents correspond to zero transfer coefficient.

It is convenient to present the system of equations for the synapse
cell in terms of Josephson junctions phase sum, $\varphi_+ =
(\varphi_1 + \varphi_2)/2$, and phase difference, $\varphi_- =
(\varphi_1 - \varphi_2)/2$:
\begin{subequations} \label{seq1}
\begin{equation}
\varphi_+ + \left(\frac{l}{2} + l_{in}\right)i_{in} + \phi_{in} = 0,
\end{equation}
\begin{equation}
\varphi_- + l i_{cir} = 0.
\end{equation}
\end{subequations}
Further, introducing the sum $\Sigma i_c = i_{c1} + i_{c2}$ and
difference $\Delta i_c = i_{c1} - i_{c2}$ of the critical currents,
and taking (\ref{CPR}) into account one can represent (\ref{seq1})
in the following form:
\begin{subequations} \label{seq2}
\begin{multline}
\varphi_+ + \left(\frac{l}{2} + l_{in}\right)(\Sigma i_c
\sin\varphi_+\cos\varphi_- + \Delta i_c \sin\varphi_-\cos\varphi_+)
\\ + \phi_{in} = 0,
\end{multline}
\begin{equation}
\varphi_- + \frac{l}{2} (\Sigma i_c \sin\varphi_-\cos\varphi_+ +
\Delta i_c \sin\varphi_+\cos\varphi_-) = 0.
\end{equation}
\end{subequations}
The dependence of the phase difference on the phase sum,
$\varphi_-(\varphi_+)$, can be obtained \cite{SUSTSQ,JETPSQ} from
(\ref{seq2}b) with corresponding function
\begin{multline}
f(\varphi_-,\varphi_+) = \varphi_- \\ + \frac{l}{2} (\Sigma i_c
\sin\varphi_-\cos\varphi_+ + \Delta i_c \sin\varphi_+\cos\varphi_-)
\end{multline}
as follows
\begin{equation} \label{f-f+}
\varphi_- = \int_0^{\pi \sgn\Delta i_c} \Hev[-f(x,\varphi_+) \sgn
\Delta i_c]dx,
\end{equation}
where $\Hev(x)$ is the Heaviside step function. Equations
(\ref{seq1}b), (\ref{seq2}a) and (\ref{f-f+}) implicitly define the
cell transfer function $\phi_{out}(\phi_{in})$ through dependencies
$\phi_{out} = 2li_{cir} =  - 2\varphi_-(\varphi_+)$ and
$\phi_{in}[\varphi_-(\varphi_+),\varphi_+]$ on $\varphi_+$. Here we
are interested in the range of the phase sum, $\varphi_+ \in
[0,\pi/2)$, where the transfer function might be linear.

Figure~\ref{Fig2}b shows synapse cell transfer function for
different values of critical currents difference in the range
$\Delta i_c \in [-0.9,0.9]$. The critical current sum is $\Sigma i_c
= 1$. With the fixed critical currents the shape of the transfer
function is determined by inductances $l_{in}$, $l$.

In accordance with (\ref{seq1}a), an increase in input inductance
$l_{in}$ increases the amplitude of nonlinearity of the dependence
of input current on input flux $i_{in}(\phi_{in})$ making it more
tilted. This is in complete analogy with parametric quantron scheme
(\ref{PQeq}). The slope of the linear part of the transfer function
is correspondingly decreased. However, this gives a stretching of
this linear part, which is of use for us, and contraction of the
nonlinear part.

Increase in inductance $l$ provides the same effect (see
(\ref{seq1}a)). At the same time it increases the nonlinearity of
the dependence of output flux on phase sum (see (\ref{seq2}b)) which
vice versa increases the slope of the linear part though making it
less linear. The goal of optimization of the transfer function
$\phi_{out}(\phi_{in})$ is the maximum modulation of its slope
alongside with the high linearity among the possibly wider range of
input flux.

In our case the values of inductances were chosen to be $l_{in} =
2$, $l = 4$. With these parameters magnetic flux can be transferred
through the synapse with coefficients in the range $\sim (-0.5,0.5)$
depending on the critical currents difference. For maximum output
magnetic flux of optimized neuron, $2\pi l_{out}/(l_a + l_{out})
\approx 1.1$, maximum standard deviation of the synapse transfer
function from the linear function is at the level of $10^{-3}$. In
the whole shown range $[0,\pi]$ it is of an order of magnitude
worse.

\section{Discussion}

Both considered cells operate in a pure superconducting regime.
Evolution of their states is fully physically reversible. Therefore,
they can be operated adiabatically with energy per operation down to
the Landauer limit \cite{Land}. For standard working temperature of
superconducting circuits, $T = 4.2$~K, this limit corresponds to the
energy, $k_B T \ln 2 \approx 4 \times 10^{-23}$~J (where $k_B$ is
the Boltzmann constant). Estimations show that the bit energy can be
as low as $10^{-21}$~J for adiabatic superconductor logic at clock
frequency of 10 GHz \cite{EEASL}. This is million times less than
characteristic energy consumed by a semiconductor transistor. In one
hand, with taking into account the fact that modern implementation
of neuron based on complementary-metal-oxide semiconductor (CMOS)
technology requires a few dozens of transistors the possible gap
between power consumption of semiconductor and superconductor ANN is
increased by an order. On the other hand, penalty for
superconducting circuits cooling is typically several hundreds W/W
that cancels out the two to three orders of supremacy. Nevertheless,
the proposed adiabatic superconducting ANN can be up to
$10^4~-~10^5$ times more energy efficient than its semiconductor
counterparts.

One should note some peculiarities of the proposed concept. First of
all, there is no power supply in these circuits and so the signal
vanishes. Therefore, there is a need for a flux amplifier which can
be implemented on a base of some standard adiabatic cell like
adiabatic quantum flux parametron (AQFP) \cite{Beil,AQFP}. However,
such aspects as the linearity of amplification, the distance of
signal propagation without amplification and related issues of
achievable fan-in and fan-out should be additionally considered.

Another feature is the periodicity of sigma-cell based neuron
transfer function. Corresponding issues can be mitigated by a signal
normalization.

Along with the using of standard superconducting integrated circuits
fabrication process, the proposed cells require utilization of
magnetic Josephson junctions which are relatively new to
superconducting technology. Nevertheless, modern developments of
cryogenic magnetic memory \cite{Beil,SMRAM} and superconducting
logic circuits with controlled functionality \cite{SFPGA} promise
their fast introduction.

In particular case of the proposed synapse, one could benefit from
implementation of magnetic Josephson junction controlled by
direction of magnetic field, like Josephson magnetic rotary valve
\cite{RotValve} with heterogeneous area of weak link. Such valve is
featured by high critical current for a certain direction of its
F-layer magnetization and low critical current for the direction
rotated by 90 degrees. Two such junctions in close proximity to each
other with mutual rotation on 90 degrees relative to their axes
directed along the boundary of inhomogeneity allows to obtain high
critical current for one junction and low critical current for
another one with the same direction of magnetizations of their
F-layers. In this case, rotation of their magnetizations leads to
corresponding decrease and increase of Josephson junctions critical
currents which means modulation of synapse weight, according to
Fig.~\ref{Fig2}. Utilization of the rotary valve reduces the number
of control lines required to program the magnetic Josephson
junctions by half. However, their total number, which is twice the
number of synapses, remains huge for practical ANNs. Therefore, the
effective synapse control is another urgent task on the way to
multilayer adiabatic superconducting ANN.

\section{Conclusion}

In this paper, we considered operation principles of adiabatic
superconducting basic cells for implementation of multilayer
perceptron. These are artificial neuron and synapse which are
nonlinear and close-to-linear superconducting transformers of
magnetic flux, respectively. Both cells are capable of operation in
adiabatic regime featured by ultra-low power consumption at the
level of 4 to 5 orders of magnitude less than that of their modern
semiconductor counterparts (including cooling power penalty). The
proposed neuron cell contains just a single Josephson junction. The
neuron provides one-shot calculation of either sigmoid or hyperbolic
tangent activation function. The certain type of this function is
determined by the type of utilized Josephson junction and can be
also switched on the fly by application of magnetic flux. The
synapse is implemented with two magnetic Josephson junctions with
controllable critical currents. It provides both positive and
negative signal transfer coefficients in the range $\sim
(-0.5,0.5)$. The presented concept of adiabatic superconducting
neuromorphic circuits promises to be a compact and the most energy
efficient solution for the artificial neural network of considered
type.


\section{Acknowledgments}
This work was supported by grant No. 17-12-01079 of the Russian
Science Foundation. A.E.S. acknowledges the Basis Foundation
scholarship.

\end{document}